\documentclass[prb,twocolumn,showpacs,superscriptaddress]{revtex4-1}
\usepackage{graphicx}
\begin{document}
\title{Disorder effects and current percolation in FeAs based superconductors}

\author{M~Eisterer}
\author{M~Zehetmayer}
\author{H~W~Weber}

\address{Atominstitut, Vienna University of Technology, 1020 Vienna, Austria}

\author{J~Jiang}
\author{J~D~Weiss}
\author{A~Yamamoto}
%\author{A~A~Polyanskii}
\author{E~E~Hellstrom}
\author{D~C~Larbalestier}

\address{National High Magnetic Field Laboratory, Florida State University, Tallahassee,
FL 32310, USA }

\author{N~D~Zhigadlo}
\author{J~Karpinski}

\address{Laboratory for Solid State Physics, ETH Zurich, CH-8093 Zurich, Switzerland
}

%\ead{eisterer@ati.ac.at}

\begin{abstract}
We report the influence of atomic disorder introduced by
sequential neutron irradiation on the basic superconducting
properties, flux pinning and grain connectivity. Two different
polycrystalline Sm-1111 samples (SmFeAsO$_{1-x}$F$_{x}$) and two
Ba-122 single crystals (BaFe$_{1.8}$Co$_{0.2}$As$_2$) were
investigated. The monotonous decrease of the transition
temperature with neutron fluence degrades the upper critical
field, at least in the investigated temperature region. Pinning on
the other hand is largely improved, with a different optimal
defect concentration (fluence) in the two materials. The analysis
of the current flow in the polycrystalline samples reveals weak
link behaviour in the majority of grain connections and the
existence of stronger grain connections. The density of the latter
seems to be close to the percolation threshold (i.e. the minimum
fraction for a continuous current path). Both types of connections
are sensitive to disorder and degrade upon neutron irradiation.
\end{abstract}

%\pacs{74.70.Xa, 74.62.En, 74.81.Bd, 74.25 Sv, 61.80.Hg, 74.25.Op, 74.25.Wy}
% pnictides, disorder, granular sc, Ic, neutron irr., Bc, pinning

\maketitle

\newpage

\section{Introduction}

The optimization of the superconducting properties, flux pinning,
and intergranular current flow are important issues for
applications of superconductors, which can be addressed by neutron
irradiation studies
\cite{Mei85,Zeh04,Put05,Put06,Kru07,Wil06,Sau98,Eis09b,Oss95}.
Fast neutrons introduce atomic disorder and in many cases pinning
efficient defects. The enhanced impurity scattering potentially
changes the transition temperature
\cite{Put05,Put06,Kru07,Wil06,Sau98,Eis09b,Oss95}, the
superconducting energy gap \cite{Put06}, the upper critical field
\cite{Put05,Kru07,Oss95}, and the magnetic penetration depth
\cite{Oss95}. These defects are a model system for naturally grown
or artificially produced defects \cite{Kru07,Eis07} and help to
understand sample to sample variations and to identify possible
routes for material optimization \cite{Eis05}. Theoretical
predictions can be checked \cite{Put06} and limitations of the
current flow identified \cite{Ton01,Eis05}, since inter- and
intragranular defects are differently affected. The influence of
nanosized defects as bulk pinning centers was unambiguously
demonstrated \cite{Mei85,Sau98,Zeh04}.

Fe-based superconductors offer promising superconducting
properties, i.e. comparatively high transition temperatures
\cite{Ren08b,Ren08c,Yan08,Rot08} and high upper critical fields
\cite{Jar08,Sen08b,Yam09}. The intergranular current flow,
however, seems to be problematic
\cite{Yam08,Kam09,Kam09b,Lee09,Ota09,Tam09}. Very few irradiation studies on bulk samples \cite{Eis09b, Kar09, Moo09} and single crystals \cite{Eis09c} were performed until now.

In this paper we report the influence of neutron irradiation on
the transition temperature, the upper critical field, and flux
pinning. The inter- and intragranular current flow is
investigated.

\section{Experimental\label{secexp}}

One SmFeAsO$_\mathrm{1-x}$F$_\mathrm{x}$ sample (denoted as
Sm-1111A in the following) was prepared at the National High
Magnetic Field Laboratory. The starting materials of As, Sm, Fe,
Fe$_2$O$_3$ and SmF$_3$ were mixed and pressed into a pellet,
wrapped with Nb foil, and sealed in a stainless steel tube. The
sealed sample was heat treated at 1160\,$^\circ$C for 6 hours in a
high temperature isostatic press under a pressure of 280\,MPa. The
main phase of the sample is SmFeAsO$_\mathrm{1-x}$F$_\mathrm{x}$,
with a grain size of 10 to 15 micrometers. The impurity phases
include SmAs, SmOF and FeAs. The size of the sample was $1.1\times
1.7 \times 3.7$\,mm$^3$.

The second polycrystalline sample of
SmFeAsO$_\mathrm{1-x}$F$_\mathrm{x}$ (sample Sm-1111B) was
synthesized at the ETH Zurich \cite{Zhi08}. SmAs, FeAs, SmF$_3$,
Fe$_2$O$_3$, and Fe were pulverized and sealed in a BN crucible.
At a pressure of 3\,GPa, the temperature was increased to
1350–-1450\,$^\circ$C within 1 h. After keeping this temperature
for 4.5 h, the crucible was quickly cooled down to room
temperature. The sample investigated in this study was a piece of
the pellet with an approximately triangular cross section ($\sim
2.5$\,mm$^2$) and a height of 1.9\,mm.

BaFe$_{1.8}$Co$_{0.2}$As$_2$ single crystals were prepared by the
self-flux method \cite{Sef08} under a pressure of 280\,MPa at the
National High Magnetic Field Laboratory. One crystal was
characterized resistively, another inductively. Typical dimensions
of these crystals were $1.4\times0.7\times0.1$\,mm$^3$.

Neutron irradiation was performed in the central irradiation
facility of the TRIGA-Mark-II reactor in Vienna. The samples were
sealed into a quartz tube and sequentially irradiated, starting
with a fast neutron fluence ($E>0.1$\,MeV) of $4\times
10^{21}$\,m$^{-2}$. The highest cumulative fluence was $1.8\times
10^{22}$\,m$^{-2}$.

Neutrons transfer their energy to the lattice atoms by direct
collisions. The transferred energy must exceed the binding energy
of the lattice atom to displace it, thus only fast neutrons lead
to defects. The size of defects is expected to range from point
defects to several nm, as observed in the cuprates \cite{Fri93}. 
Self shielding effects can be neglected, since the penetration depth of fast neutrons is of the order of a few centimeters, which is much larger than the sample
dimensions. Only neutrons of low or intermediate energies are
shielded efficiently because of large neutron cross sections at
these energies. Some of these neutron capture reactions are
followed by $\gamma$- or $\beta$-emissions, which may produce
single displaced atoms \cite{Eis09c}. These reactions are restricted to the surface in Sm-1111, since Sm is a paricular strong neutron absorber. The bulk will be penetrated only by fast neutrons. In contrast, the recoil from $\gamma$- and $\beta$-emissions might also contribute to the total defect density in the bulk of Ba-122. However, since we do not find any significant difference in the change of the transition temperature in the two compounds after irradiation, defects seem to result predominantly from fast neutrons.      

The resistivities of sample Sm-1111A and of one Ba-122 crystal
were measured at various fixed fields while cooling at a rate of
10\,K/h with an applied current of 10\,mA (single crystal:
300\,$\mu$A). Current and voltage contacts were made by silver
paste (single crystal: silver epoxy). The distance between the two
voltage contacts was around 1\,mm for the polycrystalline samples
and around 0.3\,mm for the crystal. The transition temperature and
the upper critical field were determined by means of a 90\%
criterion.

Magnetization loops at various temperatures were recorded in a
commercial 7\,T SQUID magnetometer (sample Sm-1111A) and a
commercial vibrating sample magnetometer (Ba-122). The critical
current density was calculated from the magnetization loops using
the Bean model. A self field correction was made for the single
crystal. The self field was calculated numerically, averaged over the sample volume, and added to the applied field $\mu_0H$, leading to $J_\mathrm{c}(B)$\cite{Wie92}. $J_\mathrm{c}$ of sample Sm-1111A
represents a rough estimation because of the uncertainty in the
geometry of the current loops (grain size). The self field
correction was therefore abandoned ($J_\mathrm{c}(\mu_0H)$).

The ac susceptibility at 33\,Hz was measured in the same SQUID
with an amplitude of 30\,$\mu$T (Ba-122 single crystal:
100\,$\mu$T) to determine $T_\mathrm{c}$ and the shielding
fraction. The demagnetization factor, which is needed for
estimating the shielding fraction, was calculated numerically for
the actual sample geometry.

In addition, the remnant magnetic moment was measured (SQUID) as
function of the maximally applied field, $H_\mathrm{max}$, in
order to check for magnetic granularity
\cite{Mul94,Ton02,Yam08,Ota09}.

\section{Results and Discussion}

\begin{figure} \centering \includegraphics[clip,width=0.45\textwidth]{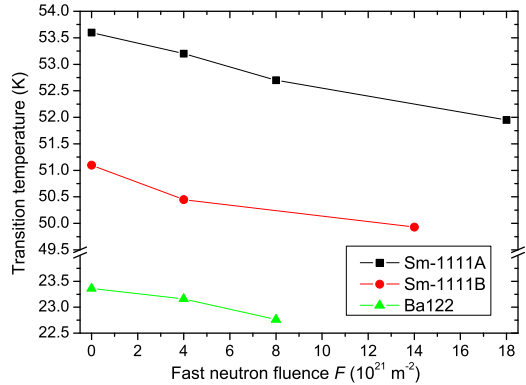}
\caption{The transition temperatures of three samples as a
function of the fast neutron fluence, $F$. The introduced disorder
decreases $T_\mathrm{c}$ of all samples with a similar slope
$\mathrm{d}T_\mathrm{c}/\mathrm{d}F$.} \label{FigTc}
\end{figure}

Neutron irradiation introduces scattering centers, which increase
the normal state resistivity. The normalized resistivity
$\rho_\mathrm{norm}:=\rho(55\,\mathrm{K})/(\rho(300\,\mathrm{K})-\rho(55\,\mathrm{K}))$
increased from 0.49 in the unirradiated sample Sm-1111A to 0.55,
0.61, and 0.72 after irradiation to a fluence, $F$, of 4, 8, and
$18\times10^{21}$\,m$^{-2}$, respectively. The normalization to
the phonon contribution
($\rho(300\,\mathrm{K})-\rho(55\,\mathrm{K})=450\pm150\,\mu\Omega$cm)
potentially cancels a reduction of the effective current carrying
cross section due to secondary phases, voids, or cracks
\cite{Row03} and errors in the determination of the distance
between the voltage taps. The nearly linear increase of
$\rho_\mathrm{norm}$ with neutron fluence ($\sim
1.3\times10^{-23}$\,m$^{2}$) indicates that the defect density
scales with neutron fluence and that the resistivity is still far
from saturation, where the mean free path of the charge carriers
approaches the lattice parameter. The increase of impurity
scattering is also evidenced by the resistivity ratio
$\rho(300\,\mathrm{K})/\rho(55\,\mathrm{K})$, which decreases from
3.03 to 2.83, 2.64, and 2.38.

This resistivity ratio was only 2.2 in the Ba-122 single crystal
and decreased to 1.9 after irradiation to
$4\times10^{21}$\,m$^{-2}$. A clear increase of resistivity after
irradiation is thus observed in both compounds, in single and
polycrystals.

Neutron irradiation also decreases the transition temperature
(Fig.~\ref{FigTc}) with a similar slope in both materials:
$\mathrm{d}T_\mathrm{c}/\mathrm{d}F=-9.1$, $-7.75$, and
$-7.5\times10^{-23}$\,K\,m$^{2}$ in Sm-1111A (resistive), Sm-1111B
(inductive), and Ba-122 (inductive), respectively. This decrease
of the transition temperature caused by impurity scattering is a
general property of superconductors with anisotropic energy gap
\cite{Mil88}, and particularly with s$_\pm$ (extended s-wave
\cite{Maz08}) gap symmetry \cite{Ban09}, or a consequence of
multiband superconductivity \cite{Put06}.

Karkin et al \cite{Kar09} found a complete suppression of superconductivity in La-1111 after irradiation to a fast neutron fluence of $1.6\times10^{23}$\,m$^{-2}$, which is about one order of magnitude higher than the highest fluence of the present study. The slope $\mathrm{d}T_\mathrm{c}/\mathrm{d}F$ found in our samples would predict a decrease in $T_\mathrm{c}$ of between 12\,K and 15\,K at their neutron fluence. Either the decrease of $T_\mathrm{c}$ with neutron fluence does not stay linear at high fluences, or, their sample behaves differently. Note that $T_\mathrm{c}$ of the La-1111 sample was significantly more below optimum than that of our Sm-1111 samples.  

\begin{figure} \centering \includegraphics[clip,width=0.45\textwidth]{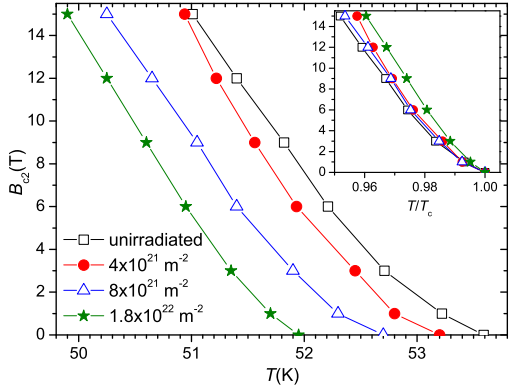}
\caption{Temperature dependence of the upper critical field,
$B_\mathrm{c2}$, in sample Sm-1111A. The insert contain the same
data, but the temperature was normalized by the transition
temperature, $T_\mathrm{c}$.  The influence of neutron irradiation
is dominated by the decrease in transition temperature.
Nevertheless, the slope increases after neutron irradiation,
having its maximum after the first irradiation step
($4\times10^{21}$\,m$^{-2}$).} \label{FigHc2}
\end{figure}

The upper critical field of sample Sm-1111A near the transition
temperature is presented in Fig.~\ref{FigHc2}. The decrease in
transition temperature shifts the $B_\mathrm{c2}(T)$ curves to
lower temperatures, but the slope increases after irradiation (see
the insert). The interplay between the decrease in transition
temperature and the increase in slope suggests the presence of a
maximum in $B_\mathrm{c2}(0)$ at around
$4\times10^{21}$\,m$^{-2}$.

In the Ba-122 system, the slope of $B_\mathrm{c2}(T)$ hardly
changes after irradiation for $H\parallel ab$ leading to a general
decrease of the upper critical field in this field orientation.
$B_\mathrm{c2}(T)$ becomes steeper for $H\parallel c$, which
results in a reduction of the upper critical field anisotropy (e.g. from 2.8 to 2.5 at 22 K)\cite{Eis09c}. Good agreement of the angular dependence of
$B_\mathrm{c2}$ with anisotropic Ginzburg-Landau theory is found
near $T\mathrm{c}$ before and after irradiation.

\subsection{Flux pinning\label{SecFluxpinning}}

\begin{figure} \centering \includegraphics[clip,width=0.45\textwidth]{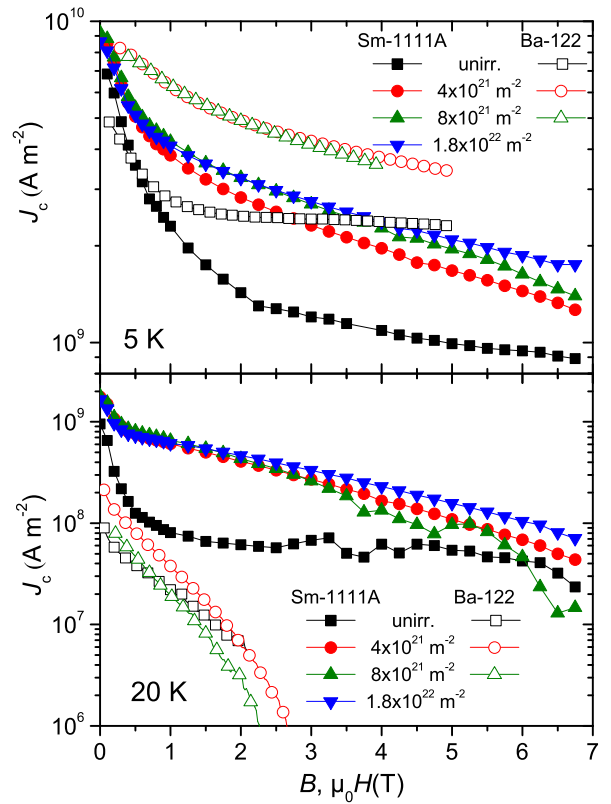}
\caption{The critical current density increases in Sm-1111 (sample
A, solid symbols) and Ba-122 (open symbols) due to the
introduction of pinning efficient defects by neutron irradiation.
The x-axis refer to $B$ for the Ba-122 crystal and to $\mu_0H$ for
the Sm-1111 sample, since no self field correction was made in the
latter case. $J_\mathrm{c}$ is an estimate for the intragranular
currents in Sm-1111.} \label{FigJcB}
\end{figure}

The critical current densities of the polycrystalline sample
Sm-1111A and the single crystal Ba-122 are compared before and
after neutron irradiaton to various fluences in Fig.~\ref{FigJcB}.
The single crystal data refer to a well defined crystallographic
orientation of the applied field ($H\parallel c$) and the induced
currents (parallel to the $ab$ planes), whereas the magnetic
moment of the polycrystalline sample results from currents flowing
in all crystallographic directions providing only a rough estimate
of the intragranular currents (see below). However, the magnitude
and field dependence of $J_\mathrm{c}$ in both samples is similar
at 5\,K (upper panel), in particular after irradiation. Both
materials show traces of a second peak (e.g. Ba-122 at 5\,K and
Sm-1111 at 20\,K), which disappear after irradiation
\cite{Eis09b,Eis09c}. The second peak effect is generally assumed
to indicate the transition from a low $J_\mathrm{c}$ ordered
vortex phase at low magnetic fields to a high $J_\mathrm{c}$
disordered phase at higher fields. Introducing more disorder
shifts the transition field to lower values and eventually leads
to the disappearance of the ordered phase, as is observed in our
samples upon irradiation. The introduced defects are efficient
pinning centers in both materials.

The self field $J_\mathrm{c}$ increases after the first
irradiation step, but does not change much upon further
irradiation. Higher fluences decrease the field dependence of
$J_\mathrm{c}$ in Sm-1111, but not in the Ba-122 single crystal,
where the second irradiation step decreases $J_\mathrm{c}$ at
20\,K, a consequence of the reduced transition temperature.

\begin{figure} \centering \includegraphics[clip,width=0.45\textwidth]{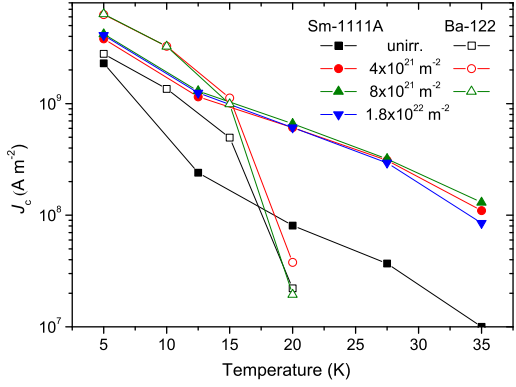}
\caption{Critical current densities at 1\,T in samples Ba-122 and
Sm-1111A as a function of temperature prior to and after neutron
irradiation to various fluences.} \label{FigJcT}
\end{figure}

The temperature dependence of the critical current densities at
1\,T is shown in Fig.~\ref{FigJcT}. The relative enhancement after
irradiation (by up to one order of magnitude) increases with
temperature (sample Sm-1111A, solid symbols), which indicates that
the pinning energy of the radiation induced defects can better
compete with the thermal energy than the pinning energy of the as
grown defect structure. A reasonable explanation is a larger
defect size, since $J_\mathrm{c}$ is not very sensitive to the
defect density (fluence).

In the Ba-122 crystal on the other hand, the temperature
dependence hardly changes after irradiation, only at high
temperatures the slope becomes slightly steeper, which is caused
by the reduced transition temperature (20\,K is close to
$T_\mathrm{c}$). The pinning energies of the as-grown and the
radiation induced defects seem to be similar.

\subsection{Intra- and intergranular currents}

\begin{figure} \centering \includegraphics[clip,width=0.45\textwidth]{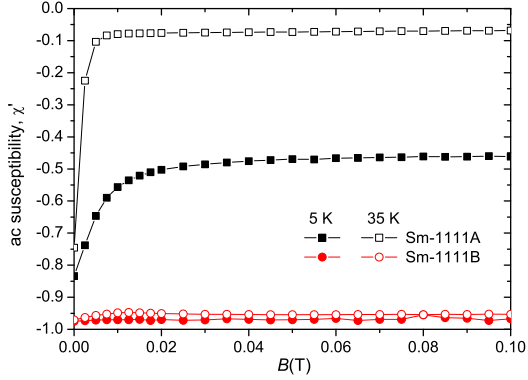}
\caption{Ac susceptibility as a function of the applied dc field.
While the global currents are rapidly suppressed in sample
Sm-1111A, they remain significant ($\gg 2.5\times
10^4$\,Am$^{-2}$) in sample Sm-1111B.} \label{FigAC}
\end{figure}

The global current flow was checked by ac susceptibility
measurements (Fig.~\ref{FigAC}). The ac susceptibility, $\chi'$,
of sample Sm-1111A rapidly increases (shielding decreases) with applied dc field, which indicates a rapid decoupling of the grains. A metallic Fe--As
wetting-phase, which can be passed by the supercurrents only via
Josephson tunneling, was found at the grain boundaries of a
comparable sample \cite{Kam09b}. These superconducting tunnel
currents are known to be rapidly suppressed by magnetic fields.

Shielding is perfect ($\chi'=-1$) within experimental accuracy in
sample Sm-1111B. This means that the supercurrents only flow at
the sample surface and the penetration depth of the applied ac
field ($\delta=H^\mathrm{ac}/J_\mathrm{c}$)is negligible compared
to the smallest sample dimension $R^s$; thus, the intergranular
currents must be \emph{much larger} than $H^\mathrm{ac}/R^s\approx
2.5\times 10^4$\,Am$^{-2}$. The field independent susceptibility
proves the existence of significant global currents, at least in
the field region relevant for the measurement of the remnant
magnetic moment. This does not necessarily mean that weak
(Josephson) grain coupling is absent in this sample. However, the
number of stronger grain connections must be significantly above
the percolation threshold \cite{Eis09}, which is defined as the
minimum fraction for a continuous current path.

\begin{figure} \centering \includegraphics[clip,width=0.4  \textwidth]{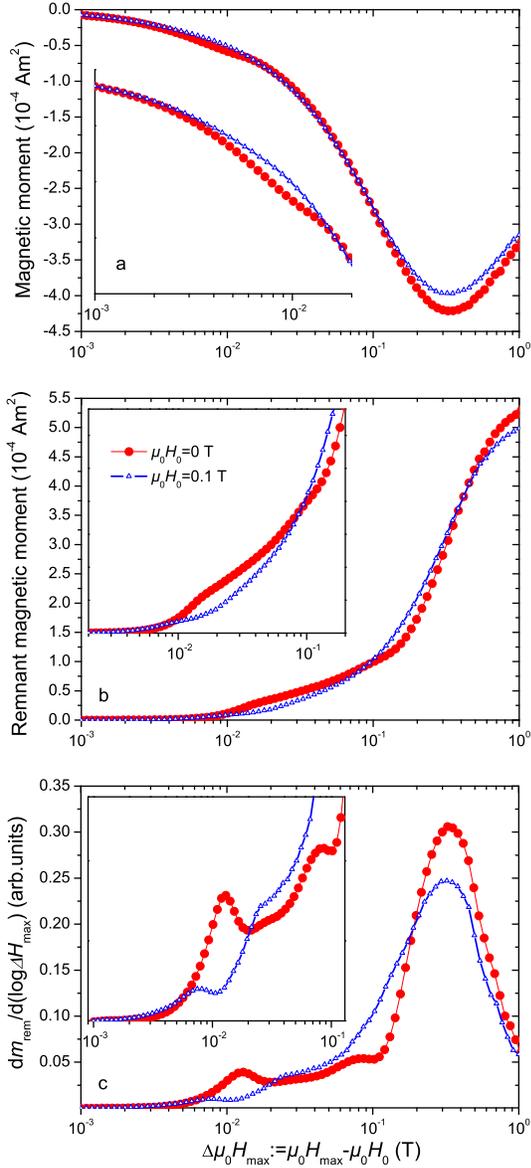}
\caption{Magnetic moment as a function of the (increasing) applied
magnetic field $\Delta\mu_0H_\mathrm{max}$ (panel a). Measurements
were done in a background field, $\mu_0H_0$, of 0\,T (circles) and
0.1\,T (triangles). The field was reduced to $\mu_0H_0$ between
each value of $\mu_0H_\mathrm{max}$ in order to measure the
remnant magnetic moment $m_\mathrm{rem}$ (panel b). A peak occurs
in $\mathrm{d}m_\mathrm{rem}/\mathrm{d}(\log\Delta
H_\mathrm{max})$, each time $\Delta H_\mathrm{max}$ approximately
equals a penetration field (panel c). The low and high field peaks
result from flux penetraion into the whole sample (intergranular
peak) and the grains (intragranular peak), respectively. All peaks
shift to lower fields at $\mu_0H_0=0.1$\,T. The shift of the
intergranular peaks is more pronounced, indicating a stronger
field dependence of $J_\mathrm{c}^\mathrm{inter}$. Data were
obtained at 5\,K on sample Sm-1111B after the first irradiation
step ($4\times 10^{21}$\,m$^{-2}$). The inserts enlarge
interesting details.} \label{FigKHMvB}
\end{figure}

In order to investigate the influence of disorder on the inter-
and intragranular currents, the remnant magnetic moment,
$m_\mathrm{rem}$, was measured as a function of the maximally
applied field, $H_\mathrm{max}$ (e.g. solid circles in
Fig.~\ref{FigKHMvB}b). The magnetic moment was measured at each
field $H_\mathrm{max}$ (Fig.~\ref{FigKHMvB}a) and after reducing
the field to zero (circles in Fig.~\ref{FigKHMvB}b). If all
current loops, which were induced by the application and removal
of the external magnetic field, are of similar geometry, a peak in
the derivative of the remnant magnetic moment with respect to the
logarithm of the maximally applied magnetic field,
\begin{equation}
\frac{\mathrm{d} m_\mathrm{rem}}{\mathrm{d}\log(H_\mathrm{max})}
\label{dmdlogH},
\end{equation}
occurs, when $H_\mathrm{max}$ approximately equals the Bean
penetration field $H^\star$ \cite{Mue99}. The critical current
density can be estimated from $H^\star \approx J_\mathrm{c}R$, if
the representative geometry of the current loops, $R$, is known
(typically the smallest sample or grain dimension).

Two peaks are observed in granular materials \cite{Mue99}, if
$J_\mathrm{c}^\mathrm{intra}R^\mathrm{g}\ll
J_\mathrm{c}^\mathrm{inter}R^\mathrm{s}$, where
$J_\mathrm{c}^\mathrm{intra}$ and $J_\mathrm{c}^\mathrm{inter}$
denote the critical current density within or between the grains,
respectively, and $R^\mathrm{s}$ ($R^\mathrm{g}$) is the smallest
sample (grain) geometry. Flux first penetrates the whole sample
along the grain boundaries, while the inner parts of the grains
are still free of vortices. If the above condition is not
fulfilled, the peaks overlap and cannot be properly distinguished.

\begin{figure} \centering \includegraphics[clip,width=0.4\textwidth]{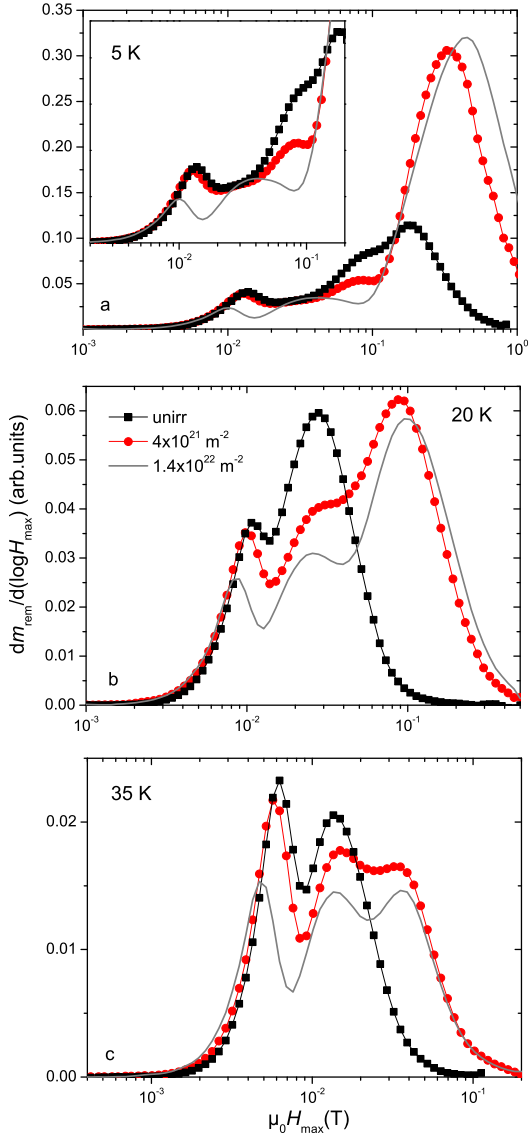}
\caption{Evolution of the peaks in
$\mathrm{d}m_\mathrm{rem}/\mathrm{d}(\log H_\mathrm{max})$ (cf.
caption in Fig.~\ref{FigKHMvB}, here with $\mu_0H_0=0$\,T) with
temperature and neutron fluence. The irradiation decreases the
intergranular currents (low and intermediate field peak). The
intragranular (high field) peak shifts to higher fields, as a
consequence of the increase in $J_\mathrm{c}^\mathrm{intra}$. The
intermediate and high field peak merge at temperatures above about
10\,K in the unirradiated sample (Sm-1111B).} \label{FigKHM}
\end{figure}

The logarithmic derivative~(\ref{dmdlogH}) of sample Sm-1111B is
presented in Fig.~\ref{FigKHM}. Two peaks occur at 20 and 35\,K
(also at 12.5 and 27.5\,K, not shown) in the unirradiated state
(solid squares). The low field peak is usually ascribed to
intergranular currents shielding the whole sample. This peak is
located at about 14\,mT at 5\,K (solid squares in
Fig.~\ref{FigKHM}a), which corresponds to a critical current
density of around $10^7$\,Am$^{-2}$ decreasing to about $5\times
10^6$\,Am$^{-2}$ at 35\,K.

It is a priori not clear whether the low field peak simply arises
from a rapid flux penetration immediately after $H_\mathrm{max}$
exceeds the lower critical field, $H_\mathrm{c1}$, of the sample,
since the reversible magnetization strongly changes around
$H_\mathrm{c1}$. In order to exclude this scenario, the
measurement was repeated after field cooling the sample at 0.1\,T.
The measurement sequence was repeated with this constant offset
field, $\mu_0H_0=0.1$\,T. The field was increased by $\Delta
H_\mathrm{max}$ and decreased to 0.1\,T, where the trapped
magnetic moment was measured. The results at 5 K are compared to
the standard measurements ($\mu_0H_0=0$\,T) in Fig.~\ref{FigKHMvB}
(the moment measured immediately after field cooling was
subtracted). The low field peak persists, but shifts to lower
fields (by a factor of 1.7), which is a result of the field
dependence of the critical currents. As can be seen in the insert
of panel~b, the remnant magnetic moment is larger (between $\Delta
H_\mathrm{max}=2$ and 8\,mT) first at 0.1\,T, which results from
an earlier flux penetration (cf. panel a). Then the larger
critical currents at 0\,T lead to a larger trapped flux, until the
same behaviour is repeated near the high field peak
($m_\mathrm{rem}$ at 0.1\,T first larger then smaller than in zero
field). In any case, the low field peak is well above the first
penetration field and obviously not caused by the reversible
magnetization, which confirms its interpretation as an
intergranular peak.

Neutron irradiation shifts the intergranular peak to lower fields
(Fig.~\ref{FigKHM}), which indicates a degradation of the
intergranular currents. The effect is small after irradiation to
$4\times 10^{21}$\,m$^{-2}$ (circles), but significant at
$1.4\times 10^{22}$\,m$^{-2}$ (line graph). Such a degradation of
the intergranular currents after neutron irradiation was also
found in Bi-tapes \cite{Ton01,Hu94} and coated conductors
\cite{Eis10}.

The high field peak usually originates from the intragranular
currents shielding only individual grains
\cite{Mue99,Ton02,Yam08}. In sample Sm-1111B, however, the high
field peak splits at 5\,K, which sheds some doubts on this
standard interpretation. A similar peak split at 5\,K was reported
for a polycrystalline Nd-1111 sample \cite{Yam08}. Three peaks are
clearly visible at all temperatures after neutron irradiation
(solid line in Fig.~\ref{FigKHM}).

It will be argued in the following, that the intermediate field
peak is caused by clusters of better connected grains and that the
high field peak is the intragranular peak. These two peaks are
only visible at 5\,K in the unirradiated sample and merge at 12.5,
20, 27.5 and 35\,K. This temperature dependence excludes the
scenario, where the two peaks are caused by currents of the same
magnitude flowing on different length scales (e.g., grains and
clusters, or grains (clusters) of different sizes). This is
confirmed by the changes after irradiation, which are quite
different for these peaks. While the high field peak shifts to
higher fields, the intermediate peak shows a less systematic
behavior. It shifts to lower fields at 5\,K and 12.5\,K but
approximately retains its position at higher temperatures
(Fig.~\ref{FigKHM}). The shift of the high field peak obviously
results from improved pinning (larger intragranular currents, see
Sec.~\ref{SecFluxpinning}) which also enhances the hysteresis
(e.g., larger saturated remnant moment). The decrease of the
intermediate peak field at low temperatures after irradiation
indicates that the corresponding currents are limited by grain
boundaries. This scenario is supported by the field dependence of
the peak positions (Fig.~\ref{FigKHMvB}). While the high field
peak shifts only by about 5\%, the low and intermediate peak
fields decrease by a factor of roughly two due to the increase of
the background field from 0 (solid circles) to 0.1\,T (open
triangles). (The intermediate field peak degrades to a bump at
0.1\,T.)

Only a rough estimate can be given for $J_\mathrm{c}$ in the
current loops which are responsible for the intermediate peak.
Since the size of the loops must be smaller than the sample
geometry and larger than the individual grains, the corresponding
current density is between $2\times 10^8$\,Am$^{-2}$ and $2\times
10^9$\,Am$^{-2}$ at 5\,K and 0\,T.

At least two types of grain connections exist in our
polycrystalline samples: (i) the wetted grain boundaries (type-A),
whose currents are rapidly suppressed by magnetic fields and (ii)
grain boundaries over which currents can pass also at high
magnetic fields (type-B). The number of type-B connections is
below the percolation threshold in sample Sm-1111A and above the
percolation threshold in sample Sm-1111B, as indicated by the ac
susceptibility. The rapid decrease of $\chi'$ directly shows the
suppression of global currents after ``switching-off'' type-A
connections by applying an external magnetic field. It was pointed
out in \cite{Kam09b} that samples of apparently similar
microstructure (probed by x-ray and SEM) showed a large variation
in the intergrain current density. It is therefore reasonable to
assume that the number of type-B connections is close to, but
below the percolation threshold in our sample Sm-1111A. In fact
the low field peak was not present in this sample, since the
global currents were too small, but we observed the intermediate
and high field peaks (at comparable fields) which might just
reflect the structure of the incipient spanning cluster at the
percolation threshold. In the ``nodes, links, and blobs'' picture,
the incipient spanning cluster is composed of blobs which are
connected by links \cite{Con82}. Removing one link decomposes the
spanning cluster, blobs are multiply connected sites (clusters of
type-B connected grains). The two peaks in sample Sm-1111A could
correspond to flux penetration into the blobs and grains
respectively. In a comparable sample which obviously contained a
spanning cluster of type-B connections, the links were made
directly visible by scanning laser microscopy \cite{Kam09b}.

In sample Sm-1111B, the density of type-B connections seems to be
significantly above the percolation threshold, since shielding of
very small ac fields was nearly perfect (see above). However, the
clusters of type-B connected grains may persist well above the
percolation threshold and be responsible for the intermediate
field peak.

\section{Conclusions}

Atomic scale disorder is not a promising way for improving the
basic superconducting properties ($B_\mathrm{c2}$, $T_\mathrm{c}$)
in the Fe-based superconductors. Bulk pinning on the other hand is
strengthened by the introduction of nanoscale defects. The
critical current densities in the grains generally increase after
the first irradiation step (fast neutron fluence: $4\times
10^{21}$\,m$^{-2}$), but start to decrease again upon further
irradiation in the Ba-122 single crystal, while a further
improvement is found at high fields in Sm-1111 up to a fluence of
$1.8\times 10^{22}$\,m$^{-2}$.

We found indications that the majority of grain connections are
weak links which decouple in magnetic fields. We identify a second
type of connections over which supercurrents flow at high fields
in a percolative manner, since their density seems to be close to
the percolation threshold.

\begin{acknowledgments}
Work at ETH Zurich was supported by the Swiss National Science
Foundation through the NCCR program MaNEP.

Work at NHMFL was supported under NSF Cooperative Agreement
DMR-0084173, by the State of Florida, and by AFOSR under grant
FA9550-06-1-0474.
\end{acknowledgments}

%\bibliographystyle{apsrev}
%\bibliographystyle{michl}
%\bibliography{ref} % ref.bib is the name of our database

\begin{thebibliography}{10}

\bibitem{Mei85}
Meier-Hirmer~R, K\"upfer~H, and Scheurer~H 1985 {\em Phys. Rev.} B
{\bf 31}
  183

\bibitem{Zeh04}
Zehetmayer~M, Eisterer~M, Jun~J, Kazakov~S~M, Karpinski~J,
Birajdar~B, Eibl~O, and Weber~H~W 2004 {\em Phys. Rev.} B {\bf 69}
054510

\bibitem{Put05}
Putti~M et~al. 2005 {\em Appl. Phys. Lett.} {\bf 86} 112503

\bibitem{Put06}
Putti~M, Affronte~M, Ferdeghini~C, Manfrinetti~P, Tarantini~C, and
  Lehmann~E 2006 {\em Phys. Rev. Lett.} {\bf 96} 077003

\bibitem{Kru07}
Krutzler~C, Zehetmayer~M, Eisterer~M, Weber~H~W, Zhigadlo~N~D, and
  Karpinski~J 2007 {\em Phys. Rev.} B {\bf 75} 224510

\bibitem{Wil06}
Wilke~R~H~T, Bud\char39{}ko~S~L, Canfield~P~C, Farmer~J, and
  Hannahs~S~T 2006 {\em Phys. Rev.} B {\bf 73} 134512

\bibitem{Sau98}
Sauerzopf~F~M 1998 {\em Phys. Rev.} B {\bf 57} 10959

\bibitem{Eis09b}
Eisterer~M, Weber~H~W, Jiang~J, Weiss~J~D, Yamamoto~A,
Polyanskii~A~A,  Hellstrom~E~E, and Larbalestier~D~C 2009 {\em
Supercond. Sci. Technol.}
  {\bf 22} 065015

\bibitem{Oss95}
Ossandon~J~G, Thompson~J~R, Kim~Y~C, Sun~Y~R, Christen~D~K, and
  Chakoumakos~B~C 1995 {\em Phys. Rev.} B {\bf 51} 8551

\bibitem{Eis07}
Eisterer~M, M\"uller~R, Sch\"oppl~R, Weber~H~W, Soltanian~S, and
  Dou~S~X 2007 {\em Supercond. Sci. Technol.} {\bf 20} 117

\bibitem{Eis05}
Eisterer~M, Krutzler~C, and Weber~H~W 2005 {\em J. Appl. Phys.}
{\bf 98}  033906

\bibitem{Ton01}
T\"onies~S, Weber~H~W, Guo~Y~C, Dou~S~X, Sawh~R, and Weinstein~R
2001 {\em Appl. Phys. Lett.} {\bf 78} 3851

\bibitem{Ren08b}
Ren~Z-An et al. 2008 {\em Chin. Phys. Lett.} {\bf 25} 2215

\bibitem{Ren08c}
Ren~Z-An et al. 2008 {\em Europhys. Lett.} {\bf
  83} 17002

\bibitem{Yan08}
Yang~J  et al. 2008 {\em Supercond. Sci. Technol.} {\bf
  21} 082001

\bibitem{Rot08}
Rotter~M, Tegel~M, and Johrendt~D 2008 {\em Phys. Rev. Lett.} {\bf
101}107006

\bibitem{Jar08}
Jaroszynski~J et~al. 2008 {\em Phys. Rev.} B {\bf 78} 064511

\bibitem{Sen08b}
Senatore~C, Fl\"ukiger~R, Cantoni~M, Gu~W, Liu~R~H, and Chen~X~H
2008 {\em Phys. Rev.} B {\bf 78} 054514

\bibitem{Yam09}
Yamamoto~A et al. 2009 {\em Appl. Phys. Lett.} {\bf 94} 062511

\bibitem{Yam08}
Yamamoto~A et~al. 2008 {\em Supercond. Sci. Technol.} {\bf 21}
095008

\bibitem{Kam09}
Kametani~F et~al. 2009 {\em Supercond. Sci. Technol.} {\bf 22}
015010

\bibitem{Kam09b}
Kametani~F et~al. 2009 {\em Appl. Phys. Lett.} {\bf 95} 142502

\bibitem{Lee09}
Lee~S et~al. 2009 {\em Appl. Phys. Lett.} {\bf 95} 212505

\bibitem{Ota09}
Otabe~E~S et~al. 2009 {\em Physica C} {\bf 469} 1940

\bibitem{Tam09}
Tamegai~T, Nakajima~Y, Tsuchiya~Y, Iyo~A, Miyazawa~K, Shirage~P~M,
  Kito~H, and Eisaki~H 2009 {\em Physica C} {\bf 469} 915

\bibitem{Kar09}
Karkin~A~E, Werner~J, Behr~G, and Goshchitskii~B~N 2009{\em Phys.
Rev.} B {\bf 80} 174512

\bibitem{Moo09}
Moore~J~D et al. 2009 {\em Supercond. Sci. Technol.}
{\bf 22} 125023

\bibitem{Eis09c}
Eisterer~M, Zehetmayer~M, Weber~H~W, Jiang~J, Weiss~J~D,
Yamamoto~A, and Hellstrom~E~E 2009 {\em Supercond. Sci. Technol.}
{\bf 22} 095011

\bibitem{Zhi08}
Zhigadlo~N~D, Katrych~S, Bukowski~Z, Weyeneth~S, Puzniak~R, and
  Karpinski~J 2008 {\em Journal of Physics: Condensed Matter} {\bf 20} 342202

\bibitem{Sef08}
Sefat~A~S, Jin~R, McGuire~M~A, Sales~B~C, Singh~D~J, and Mandrus~D
2008 {\em Phys. Rev. Lett.} {\bf 101}
  117004

\bibitem{Fri93}
Frischherz~M~C, Kirk~M~A, Zhang~J~P, and Weber~H~W 1993 {\em
Philos.
  Mag.} A {\bf 67} 1347

\bibitem{Wie92}
Wiesinger~H~P, Sauerzopf~F~M, and Weber~H~W 1992 {\em Physica C}
{\bf 203} 121


\bibitem{Mul94}
M\"uller~K-H, Andrikidis~C, Liu~H~K, and Dou~S~X 1994 {\em Phys.
Rev.} B
  {\bf 50} 10218

\bibitem{Ton02}
T\"onies~S, Vostner~A, and Weber~H~W 2002 {\em J. Appl. Phys.}
{\bf 92} 2628

\bibitem{Row03}
Rowell~J~M 2003 {\em Supercond. Sci. Technol.} {\bf 16} R17

\bibitem{Mil88}
Millis~A~J, Sachdev~S, and Varma~C~M 1988 {\em Phys. Rev.} B {\bf
37} 4975

\bibitem{Maz08}
Mazin~I~I, Singh~D~J, Johannes~M~D, and Du~M~H 2008 {\em Phys.
Rev. Lett.} {\bf 101} 057003

\bibitem{Ban09}
Bang~Y, H-Y~Choi, and Won~H 2009 {\em Phys. Rev.} B {\bf 79}
054529

\bibitem{Eis09}
Eisterer~M, Emhofer~J, Sorta~S, Zehetmayer~M, and Weber~H~W 2009
{\em Supercond. Sci. Technol.} {\bf 22} 034016

\bibitem{Mue99}
M\"uller~K-H, Andrikidis~C, Du~J, Leslie~K~E, and Foley~C~P 1999
{\em Phys. Rev.} B {\bf 60} 659

\bibitem{Hu94}
Hu~Q~Y, Weber~H~W, Sauerzopf~F~M, Schulz~G~W, Schalk~R~M,
  Neum\"{u}ller~H~W, and Dou~S~X 1994 {\em Appl. Phys. Lett.} {\bf 65}
  3008

\bibitem{Eis10}
Eisterer~M, Fuger~R, Chudy~M, Hengstberger~F, and Weber~H~W 2010
{\em Supercond. Sci. Technol.} {\bf 23} 014009 

\bibitem{Con82}
Coniglio~A 1982 {\em Journal of Physics A: Mathematical and
General} {\bf 15}
  3829

\end{thebibliography}

\end{document}